\documentclass[prl,reprint,twocolumn,preprintnumbers,amsmath,amssymb,showpacs]{revtex4-1} 

\usepackage{graphicx}
\usepackage{latexsym,amsmath,amssymb,amsfonts,amsthm,enumerate}
\usepackage{bm} 
\usepackage[T1]{fontenc}
\usepackage[english]{babel}
\usepackage[latin1]{inputenc}
\usepackage[usenames,dvipsnames]{color}

\usepackage{epsfig}
\usepackage{bm}
\usepackage[colorlinks,bookmarks=false,citecolor=blue,linkcolor=blue,urlcolor=blue]{hyperref}

\usepackage{graphicx}% Include figure files
\usepackage{dcolumn}% Align table columns on decimal point

\newcommand{\nn}{\nonumber}

\makeatletter
\newsavebox{\@brx}
\newcommand{\llangle}[1][]{\savebox{\@brx}{\(\m@th{#1\langle}\)}%
  \mathopen{\copy\@brx\kern-0.7\wd\@brx\usebox{\@brx}}}
\newcommand{\rrangle}[1][]{\savebox{\@brx}{\(\m@th{#1\rangle}\)}%
  \mathclose{\copy\@brx\kern-0.7\wd\@brx\usebox{\@brx}}}
\makeatother

\newcommand{\be}{\begin{equation}}
\newcommand{\ee}{\end{equation}}

\def\bea#1\eea{\begin{align}#1\end{align}}

\begin{document}

\title{Memory-preserving equilibration after a quantum quench in a $1d$ critical model}

\author{Spyros Sotiriadis}
\affiliation{SISSA, Via Bonomea 265, 34136 Trieste, Italy}
\affiliation{INFN, Sezione di Trieste, Italy}
\affiliation{Institut de Math\'{e}matiques de Marseille, (I2M) Aix Marseille Universit\'{e}, CNRS, 
Centrale Marseille, UMR 7373, 39, rue F. Joliot Curie, 13453, Marseille, France }
\affiliation{University of Roma Tre, Department of Mathematics and Physics, 
L.go S. L. Murialdo 1, 00146 Roma, Italy}

\date{\today}

\begin{abstract}
One of the fundamental principles of statistical physics is that only partial information about a system's state is required for its macroscopic description. This is not only true for thermal ensembles, but also for the unconventional ensemble, known as Generalized Gibbs Ensemble (GGE), that is expected to describe the relaxation of integrable systems after a quantum quench. By analytically studying the quench dynamics in a prototypical one-dimensional critical model, the massless free bosonic field theory, we find evidence of a novel type of equilibration characterized by the preservation of an enormous amount of memory of the initial state that is accessible by local measurements. In particular, we show that the equilibration retains memory of non-Gaussian initial correlations, in contrast to the case of massive free evolution which erases all such memory. The GGE in its standard form, being a Gaussian ensemble, fails to predict correctly the equilibrium values of local observables, unless the initial state is Gaussian itself. Our findings show that the equilibration of a broad class of quenches whose evolution is described by Luttinger liquid theory with an initial state that is non-Gaussian in terms of the bosonic field, is not correctly captured by the corresponding bosonic GGE, raising doubts about the validity of the latter in general one-dimensional gapless integrable systems such as the Lieb-Liniger model. We also propose that the same experiment by which the GGE was recently observed [Langen et al., Science 348 (2015) 207-211] can also be used to observe its failure, simply by starting from a non-Gaussian initial state.
\end{abstract}


\maketitle

\emph{Introduction} -- Understanding the physics of quantum many-body systems out of equilibrium is one of the most challenging open problems today \cite{revq,revq2}. Of central interest is the problem of \emph{quantum quenches} i.e. abrupt changes of the Hamiltonian parameters of a closed quantum system \cite{cc06}, especially in one-dimensional ($1d$) integrable systems where the study of quantum dynamics has led to intriguing discoveries, like the experimental observation of lack of thermalization \cite{kww06} and the theoretical prediction of the \emph{Generalized Gibbs Ensemble} (GGE) \cite{gge1} which has recently been observed experimentally \cite{expo}. The GGE is expected to describe the equilibration of local observables after a quantum quench in an integrable system by taking into account all constraints associated to its conserved charges \cite{gge1,gge2,RS,r-09,r-11,fm-10,SFM,ce-13,Mussardo13,cc06,caz06,caz09,caz10,Kehrein1,Kehrein2,KE08,CIC12,MG11,ir-11,cdeo08,cra10,bs08,mgs,fe13,cfe12,fcec13,KCI,DeNa,DWBC,SotiriadisCalabrese,bse-14,lightcone,delf14,KCC,MCKC,CSC13a,KZ15,DEng,SW15,P14,PE}. While in its standard form it was constructed exclusively out of local conserved charges \cite{cfe12,fe13}, it has been recently shown that \emph{quasi-local} charges must also be included \cite{WDNBFRC,BWFDNVC,PMWKZT,POZ1,POZ2,ANDREIa,ANDREIb,ANDREIc,ANDREId,PROSEN,Prosen2,Prosen3,Pereira,Panfil,EPC-nonlocal}. This gives an answer to the fundamental question of how much information about the initial state survives in the final values of local observables. Other aspects of interest are the asymptotic behavior towards equilibrium \cite{ce-13,fcec13,SotiriadisCalabrese,SM15,KZ15} or the \emph{restoration of symmetries} of the post-quench Hamiltonian that are absent in the initial state \cite{susy,fcec14}.

In \cite{SotiriadisCalabrese} it was shown that in the case of a quantum quench from an interacting to a free massive bosonic field theory in $1d$, the evolution eliminates memory about non-Gaussian correlations in the initial state so that the large time values of any local observable can be expressed solely in terms of the initial two-point function. This proves the validity of the GGE which for free systems is a Gaussian ensemble containing precisely the same information. The argument hinges upon the cluster decomposition property of the initial state and the presence of a non-vanishing post-quench mass. In the present work we study the case of massless evolution. We find that, unlike the massive case, the large time limit of physical observables retains memory of non-Gaussian initial correlations in the form of spatial averages of multi-point connected correlation functions (CCFs) of all orders. Therefore the equilibrium cannot be described by a Gaussian ensemble and the GGE fails, unless the initial state is Gaussian itself. Additionally, we derive exact results for the asymptotic behavior of physical observables and correlation functions at large times and distances. We find that symmetries that characterize the post-quench Hamiltonian but are in general absent in the initial state are restored by the evolution. Lastly we compare our conclusions with earlier studies discussing the scope of their validity and proposing an experimental implementation.

\emph{Model and quench protocol} -- We consider a quantum quench in a $1d$ bosonic system, starting from any massive field theory, free or interacting, to the free massless field theory. Thus the post-quench Hamiltonian is
\be
H = \int dx \; \left [ \frac12 \pi^2 + \frac12 (\partial_x\phi)^2 \right ]
\label{H}
\ee
while the pre-quench one is $H_0 = H + \int dx \; \mathcal{V}(\phi(x)) $, with $\mathcal{V}(\phi)$ some local `potential energy'. The initial state is the ground state of $H_0$ and we fix the field $\phi$ so that its initial expectation value vanishes $\langle \phi \rangle_0 = 0$ (the subscript `0' denotes the initial state). We also assume that the Hamiltonian $H_0$ is massive, meaning that its ground state exhibits short-range correlations i.e. all connected correlation functions (CCFs) decay exponentially with the distance within a finite correlation length $\xi_0=1/m_0$ where $m_0$ is the (renormalized) mass of its lightest particle \cite{ZJ}. 

The free massless Hamiltonian (\ref{H}) describes the simplest CFT, that is, the simplest $1d$ critical model \cite{CFTbook}. Its ground state exhibits logarithmic $\phi$ correlation functions $ \langle \phi(0,0) \phi(x,t) \rangle_{gs} = - \log (|x|^2-|t|^2) / (4\pi) + \text{const}$. 
Physical observables are however expressed in terms of the so-called `vertex operators' $V_a(x,t)=e^{ia\phi(x,t)}$ or derivatives of $\phi$, which in the ground state of $H$ exhibit power-law decaying correlation functions, as expected for a critical model, e.g. $\left \langle V_{a}(0,0) V_{-a}(x,t) \right \rangle_{gs} \sim ( {|x|^2-|t|^2}  )^{-{a^2}/({4\pi})} $. 
All other ground state multi-point functions can be derived from the above using Wick's theorem, since the ground state of any free Hamiltonian like $H$ is Gaussian. Wick's theorem for vertex operators reads 
\bea
& \Big \langle \prod_i e^{i a_i \phi_i} \Big \rangle 
=  e^{ - \frac12 \sum_i a_i^2 \langle\phi_i^2\rangle - \sum_{i<j} a_i a_j  \langle\phi_i\phi_j\rangle}
\label{Wick}
\eea
Excitations of $H$ correspond to massless particles with linear dispersion relation $E_k = |k|$ (to simplify notation, we have set the speed of sound $v=1$). 

Note that $H$ is symmetric under the continuous transformation $\phi\to\phi+\epsilon$. This is reflected in the fact that correlation functions of vertex operators vanish unless they satisfy the `neutrality condition', more explicitly $\left \langle \prod_i V_{a_i}(x_i,t_i) \right \rangle_{gs} = 0$ for $\sum_i a_i \neq 0$. 
In particular one-point functions with $a\neq 0$ vanish. The initial state instead, being the ground state of an arbitrary Hamiltonian, breaks in general this symmetry. Similarly, $H$ is symmetric under the discrete transformation $\phi\to-\phi$. 

\emph{Evolution of observables} -- We are interested in the time evolution of physical observables, in particular equal time correlation functions of vertex operators at different points, focusing on their asymptotic form at large times $t$ after the quench. The evolution of the one-point function can be derived most conveniently from the \emph{cumulant expansion}
\be
\left \langle V_a(x,t) \right \rangle = \exp \left[\sum_{n=1}^\infty \frac{(ia)^{n}}{n!}  {\llangle \phi^{n}(x,t)  \rrangle} \; \right] 
\label{eq:1}
\ee
where the double brackets $\llangle\dots\rrangle$ denote \emph{connected} correlation functions. We work in the Heisenberg picture, i.e. all correlation functions $\langle\dots\rangle$ refer to initial state expectation values of time-evolved operators. 
Similarly the two-point correlation function of vertex operators at different points $x_1\neq x_2$ is given by
\bea
& \left \langle V_{a}(x_1,t) V_{b}(x_2,t) \right \rangle \nn \\
& = \exp \left[ \sum_{n=1}^\infty \frac{i^n}{n!}  {\llangle[\big] (a \phi(x_1,t) + b \phi(x_2,t) )^{n}  \rrangle[\big]}\right]
\label{eq:v2pt}
\eea
The relation between CCFs and standard ones is that between joint cumulants and moments. 

From the above we see that it is sufficient to study the evolution of each CCF of $\phi$. 
The equation of motion for $\phi$ 
after the quench is simply the $1d$ wave equation in infinite space
\be
\ddot \phi(x,t) = \partial^2_x \phi(x,t) \;, \quad t>0 
\label{wave}
\ee
with two initial conditions $\phi(x,0)$ and $\dot\phi(x,0)$. 
The solution is given by \emph{d'Alembert's formula}
\be
\phi(x,t)=\frac12 (\phi(x-t,0) + \phi(x+t,0))  + \frac12  \int_{x-t}^{x+t} dx' \, \dot \phi(x',0)
\label{dAlembert}
\ee
or in terms of the Green's function
\bea
\phi(x,t) 
& =\sum_{s=0,1}\int dx'\; G^{(s)}(x-x',t)\phi^{(1-s)}(x',0)  \label{eq:phi(t)}
\eea
where $G(x,t)$ is the (retarded) Green's function for the $1d$ wave equation in infinite space, that is $G(x,t)=\tfrac{1}{2\pi}\int dk\, e^{ikx} ({\sin k t})/{k} = \frac12 \Theta(t-|x|)$ 
(we use the notation $f^{(0)}(t)\equiv f(t),f^{(1)}(t)\equiv\dot{f}(t)\equiv\partial_{t}f(t)$). Note the light-cone structure of $G(x,t)$. 
It then follows that the time evolution of CCFs is
\bea
 \llangle[\bigg] \prod_{i=1}^{n} \phi(x_i,t) \rrangle[\bigg] & = \int \prod^n_{i=1} dx'_i \; \prod^n_{i=1} \sum_{s_i=0,1}G^{(s_i)}(x_i-x'_i,t)  
\nn \\ & \times
\llangle[\bigg] \prod^n_{i=1} \phi^{(1-s_i)}(x'_i,0) \rrangle[\bigg]	\label{eq:2}
\eea
It is now clear that using the cumulant expansion and the last formula, we can directly express any correlation function of vertex operators in terms of a suitable convolution of initial CCFs of $\phi$ and the large time asymptotics is controlled by the scaling of the above convolution, that is, by the large distance behavior of initial correlations. Below we study separately the case of Gaussian and non-Gaussian initial states.

\emph{Gaussian initial states} -- In this special case, all $n$-point CCFs of $\phi$ with $n\geq 3$ vanish. This holds when 
$H_0$ is essentially non-interacting (i.e. either explicitly free, as when $\mathcal{V}(\phi) = \frac12 m^2_0 \phi^2$, or interacting that can be mapped into free). Since we have set $\langle\phi(x,0)\rangle=0$ we now have $\llangle \phi(0,0) \phi(x,0) \rrangle=\langle \phi(0,0) \phi(x,0) \rangle$, i.e. the only non-vanishing initial correlation function is the two-point function. 

The one-point function of vertex operators is 
$ \left \langle V_a(0,t) \right \rangle = e^{-\frac12 a^2 {\langle \phi^{2}(0,t)  \rangle}} $ 
and from (\ref{dAlembert}) and (\ref{eq:2}), taking into account the exponential large-distance decay of initial correlation functions, we find that the leading large-time behavior is given by 
$\langle \phi^{2}(0,t) \rangle = \frac12 t \int_{-t}^{+t} dr \big \langle \dot\phi(0,0) \dot\phi(r,0) \big \rangle + ... $ 
where the dots denote terms that increase slower with time for $t\gg \xi_0 $. The last integral tends to a finite value $\lambda_2$ as $t\to\infty$, since the initial two-point function $ \langle \dot\phi(0,0) \dot\phi(r,0) \rangle $ decays exponentially with the distance $r$. Therefore the leading large time behavior of the vertex operator one-point function is 
\be
\left \langle V_a(0,t) \right \rangle \propto e^{- \frac14 a^2 \lambda_2 t} 
\ee
where, using time translation invariance for $t<0$ and 
the pre-quench equation of motion $\ddot \phi(x,t) = \partial^2_x \phi(x,t) - \mathcal{V}'(\phi(x,t))$ 
evaluated at $t\to0^-$, we obtain 
\be
\lambda_2 = \int_{-\infty}^{+\infty} dr \big \langle \dot\phi(0,0) \dot\phi(r,0) \big \rangle = 
\int_{-\infty}^{+\infty} dr \big \langle \phi(0,0) \mathcal{V}'(\phi(r,0)) \big \rangle 
\label{lambda2}
\ee
We therefore find that the one-point function decays exponentially with a decay rate $\lambda_2$ determined for any Gaussian initial state by (\ref{lambda2}). Note that in the continuum limit $\lambda_2$ is typically an ultraviolet-divergent quantity \cite{cc06} meaning that it depends on microscopic details of the actual physical system (lattice spacing etc.). 

Similarly we calculate the two-point function of vertex operators, which is $\left \langle  e^{i a \phi(0,t)} e^{i b \phi(x,t)} \right \rangle = \exp{\left[ - \frac12 (a+b)^2 \langle\phi^2(0,t)\rangle - ab (\langle\phi(0,t)\phi(x,t)\rangle - \langle\phi^2(0,t)\rangle)  \right]}$ 
For large times $t\gg \xi_0$, $t>|x|/2$ we find $\langle \phi(0,t) \phi(x,t) \rangle - \langle \phi^2(0,t) \rangle = - \frac14 \int_{0}^{|x|} dx_1 \int_{0}^{|x|} dx_2 \big \langle \dot\phi(0,0) \dot\phi(x_2-x_1,0) \big \rangle + ...$, 
which for $|x|\gg \xi_0$ tends to $- \frac14 \lambda_2 |x| $. 
Essentially the asymptotics for $|x|,t\gg \xi_0$ can be easily calculated substituting $\big \langle \dot\phi(x_1,0) \dot\phi(x_2,0) \big \rangle \to \lambda_2 \delta(x_2-x_1)$. From the above we finally find
\be
\langle V_{a}(0,t) V_{b}(x,t) \rangle \propto e^{ - \frac14 (a+b)^2 \lambda_2 t + \frac14 a b \lambda_2 |x|}
\ee
i.e. two-point functions of vertex operators decay exponentially with time, unless $b=-a$ in which case they equilibrate to nonzero values. The large time two-point function $\langle V_{a}(0) V_{-a}(x) \rangle_{t\to\infty} \propto e^{- \frac14 a^2 \lambda_2 |x|}$ decays exponentially with the distance with a decay rate controlled by the same parameter $\lambda_2$ as the time decay.

For multi-point correlation functions we similarly find 
\be
\Big\langle \prod_i V_{a_i}(x_i,t) \Big \rangle \propto e^{ - \frac14 \lambda_2 (\sum_i a_i)^2 t + \frac14 \lambda_2 \sum_{i<j} a_i a_j  |x_i - x_j|}
\ee
in agreement with Wick's theorem (\ref{Wick}) which holds because both the initial state and its evolution are Gaussian. The above shows that all multi-point correlation functions of vertex operators decay exponentially with time, unless they satisfy the neutrality condition, that is, the massless evolution restores the symmetry under $\phi\to\phi+\epsilon$ transformations.

\emph{Non-Gaussian initial states} -- In the more general case of a non-Gaussian initial state, like the ground state of a genuinely interacting Hamiltonian $H_0$ (non-parabolic potential $\mathcal{V}(\phi)= \frac12 m^2_0 \phi^2 + \sum_{n=3}^\infty c_n \phi^n/n!$) the large time asymptotics of vertex correlation functions can be calculated following the same arguments as above, taking into account that for a massive self-interacting pre-quench Hamiltonian, all initial CCFs decay exponentially within a range of the order of the correlation length $\xi_0$. The vertex operator one-point function is given by (\ref{eq:1}) 
with $\phi$ correlation functions given by (\ref{eq:2}) where at large times we now find $\llangle[\big] \phi^n(0,t) \rrangle[\big] = {2^{1-n}} t \prod^n_{i=2} \int_{-\infty}^{+\infty} dr_i  \llangle \dot\phi(0,0) \prod^n_{i=2} \dot\phi(r_i,0) \rrangle + ... $ and, 
as before, for $t\gg \xi_0$ we can replace $ \llangle \prod^n_{i=1} \dot\phi(x_i,0) \rrangle \to \lambda_n \prod^n_{i=2} \delta(x_n-x_1)$ where 
\bea
\lambda_n  = \int_{-\infty}^{+\infty} \prod_{i=2}^{n} dr_i  \llangle[\Big] \dot\phi(0,0) \prod_{i=2}^{n} \dot\phi(r_i,0)  \rrangle[\Big]	\label{lambda_nG}
\eea
Thus we finally find
\bea
& \left \langle V_a(0,t) \right \rangle \propto \exp\left [ -  \left ( \sum_{n=2}^\infty \frac{(-i^{n})}{n!2^{n-1}} a^n \lambda_n \right ) t \right ]  \label{v_nG}
\eea
The expression in round brackets is a series of spatial averages of all initial multi-point CCFs. Since all $\lambda_n$ are real numbers, it is generally a complex number, whose real/imaginary part is the sum of even/odd terms respectively. This means that the one-point function decays exponentially with time with decay rate $\gamma(a)= \sum_{n=1}^\infty (-1)^{n+1} a^{2n} \lambda_{2n} /[{(2n)!2^{2n-1}}] $ 
and also exhibits oscillations with frequency $\omega(a)=\sum_{n=1}^\infty {(-1)^{n}} a^{2n+1} \lambda_{2n+1}/[{(2n+1)!2^{2n}}] $. 
If the pre-quench Hamiltonian is symmetric under reflections $\phi \to -\phi$, then all odd correlation functions vanish and the evolution is purely decaying. The quantities $\lambda_n$ clearly depend on all parameters of $H_0$: e.g. if perturbation theory describes correctly its ground state, then at first order in the couplings $c_n$ they are $\lambda_2\propto m_0$ and $\lambda_{n\geq 3} \propto c_n$. 

Similarly for the two-point function, from (\ref{eq:v2pt}) and (\ref{eq:2}) 
we find
\bea
& \left \langle V_{a}(0,t) V_{b}(x,t) \right \rangle \propto \exp \Big[ - \left(-\sum_{n=2}^\infty (a+b)^n \frac{i^{n}}{n!2^{n-1}} \lambda_n \right) t \nn \\
&- \left(\sum_{n=2}^\infty [(a+b)^n - (a^n+b^n)] \frac{i^{n}}{n!2^{n}}  \lambda_n \right) |x| \Big] \label{vv_nG}
\eea
which decays with time, unless $b=-a$ in which case it equilibrates to $\left \langle V_{a}(0) V_{-a}(x) \right \rangle_{t\to\infty} \propto e^{ - \gamma(a)|x|} $. Note that the equilibrium values depend not only on the initial two-point CCF of $\phi$ as for Gaussian initial states, but also on all even higher ones, through the values of the independent parameters $\lambda_{2n}$. This shows that information about non-Gaussian features of initial correlations survive at $t\to\infty$. It should also be noted that the evolution restores the symmetries under both $\phi\to\phi+\epsilon$ and $\phi\to-\phi$.

\emph{Failure of the bosonic Gaussian GGE} -- The GGE density matrix in one of its standard forms for free systems, is constructed using as integrals of motion the number operators $n_k$, i.e. it is written as $\rho_{\text{gge}} \propto e^{-\sum_k \beta_k n_k}$ 
with $\beta_k$ fixed by the condition that the values of the charges in the GGE equal their initial values $\langle n_k \rangle_{\text{gge}} = \langle n_k \rangle_0$ \cite{cc06}. The operators $n_k$ are linear combinations of the local integrals of motion and, in any finite lattice system, equal in number to them \cite{cfe12,fe13}. Written in this form, despite being non-local, the GGE takes into account the quasilocal charges of continuous models defined in \cite{Panfil}. 

Since this is a Gaussian density matrix, the GGE prediction for any multi-point correlation function of vertex operators satisfying the neutrality condition is given by Wick's theorem (\ref{Wick}) $\left \langle \prod_i e^{i a_i \phi(x_i)} \right \rangle_{\text{gge}} =  e^{{ - \sum_{i<j} a_i a_j \left( \langle\phi(x_i)\phi(x_j)\rangle_{\text{gge}} - \langle\phi(0)^2\rangle_{\text{gge}} \right) }}$ with $\sum_i a_i = 0$ (as already mentioned, if the neutrality condition is not satisfied the correlation functions vanish at $t\to\infty$). 
This means that all information about the GGE is included in its two-point function of $\phi$, which can be expressed in terms of $\langle n_k \rangle_0$ as $\langle\phi(0)\phi(x)\rangle_{\text{gge}} 
=\frac{1}{2\pi}\int{dk}\,{e^{ikx}}{\left[1+\langle (n_{k} +n_{-k}) \rangle_{0}\right]}/({2E_{k}}) $ where $E_{k} = |k|$. In terms of initial correlation functions, using the pre-quench equation of motion, we find
\bea
\langle\phi(0)\phi(x)\rangle_{\text{gge}} = \langle\phi(0)\phi(x)\rangle_{0} - 
\int dr \; \frac{|x-r|}{4} {\big \langle \phi(0) \mathcal{V}'(\phi(r)) \big \rangle_0 }
\eea
which can be seen as a GGE version of the \emph{virial theorem} that relates the temperature of a thermal ensemble with the inter-particle potential energy. The asymptotic form at large $|x|$ is  $\langle\phi(0)\phi(x)\rangle_{\text{gge}} \sim  - \frac14 \lambda_2 |x| + \text{const.}$, where $\lambda_2$ is given by (\ref{lambda2}). Note that in the special case $\mathcal{V}(\phi) = \frac12 m^2_0 \phi^2$ which corresponds to a free massive pre-quench Hamiltonian, the parameter $\lambda_2$ equals $m_0/2$ and the GGE is in agreement with our $t\to\infty$ results for Gaussian initial states.

As already mentioned, the above ensemble is by construction Gaussian and captures only information about the initial two-point function of $\phi$ but not of higher order initial CCFs. Therefore its predictions do not give the correct $t\to\infty$ values of vertex operators. This is also true for the generalization of the GGE that includes the quasi-local charges of \cite{Panfil} since that too is Gaussian. The fact that local observables retain memory of all even initial CCFs suggests that an ensemble that describes correctly the $t\to\infty$ limit would also include all products of $n_k$ as conserved quantities
\be
\rho_{DE} = Z^{-1} \exp\Bigg(-\sum_{\ell=1}^\infty \sum_{k_1,...,k_\ell} \beta_{k_1,...,k_\ell} \prod_{j=1}^\ell n_{k_j}\Bigg)
\label{DE}
\ee
This is clearly equivalent to the Diagonal Ensemble, i.e. the expansion in terms of projections onto all energy eigenstates \cite{gge2}. This ensemble includes \emph{all} information about the initial state that contributes to the infinite time average of \emph{any} observable in a \emph{general} system \cite{gge2,revq2} and is therefore both non-economic and impractical. It is still possible that a suitable non-Gaussian generalization of $\rho_{\text{gge}}$ that would include extra conserved charges that are special of CFT can describe correctly the large time limit. In either case, the free massless $1d$ field theory apparently retains the maximum possible memory of its initial state. 

Note that, while it has been very early pointed out that the GGE does not capture information about initial correlations between different $n_k$ \cite{GP08,KE08,caz09}, in the thermodynamic and large time limit \emph{local} observables typically lose memory of such correlations \cite{caz09,RS14}, which is the deeper reason that allows their economic description through the GGE. In contrast, in the present problem, information about all such correlations survives in the above limit and remains accessible through local measurements. To the best of our knowledge this is the first example of its kind. An intuitive explanation of this memory effect is that, due to the ballistic nature of the evolution, entangled clusters of quasiparticles carry the initial correlations intact up to infinite time. 

\emph{Discussion of results} -- The massless Gaussian field theory described by (\ref{H}) is a prototypical model of $1d$ physical systems with gapless (phononic) excitations. Indeed the `bosonization' method shows that the low-energy behavior of many interacting systems, both fermionic or bosonic, can be described by the harmonic-fluid or Luttinger liquid (LL) theory \cite{Haldane,Giamarchi,Caz04}. These include the Lieb-Liniger model (which is supposed to describe the Quantum Newton's Cradle experiment \cite{kww06}) and the gapless phase of many lattice models (e.g. Bose-Hubbard model in the superfluid phase) and spin chains (e.g. the gapless phase of the XXZ model) in the scaling limit. The Hamiltonian of the Luttinger model is $ H_{LL} = \frac{1}{2} v \int dx  \left( \frac{K}{\pi} (\partial_x \theta)^2 + \frac{\pi}{K} n^2 \right) $ 
where the density $n(x)$ and the phase field $\theta(x)$ obey canonical commutation relations and the parameters $v$ and $K$ are the sound velocity and Luttinger parameter respectively. The above form is equivalent to (\ref{H}) as can be seen by rescaling the field and the space-time coordinates.

In this context our findings show that the GGE as interpreted in the LL approximation, i.e. constructed out of the local charges of the bosonic density-phase fields, does not describe correctly the steady state corresponding to a general initial state. We emphasize that LL theory is only a low-energy approximation of the original models, while a quantum quench may create arbitrarily high-energy excitations. Therefore in this context our results do not necessarily mean that the GGE conjecture itself is incorrect: it is possible that the GGE constructed out of the exact conserved charges of the original model is correct. In particular in the case of an instantaneous interaction quench in the Lieb-Liniger model, the presence of high-energy excitations \cite{KCI,DWBC} makes the LL approximation non-applicable and at least the special cases studied in the literature seem to suggest that the GGE expressed in terms of the exact Lieb-Liniger conserved charges is correct \cite{DWBC,MC12,KCC,P14,PE,SotiriadisCalabrese}. Nevertheless the LL approach turns out to describe correctly the quench dynamics in several other cases \cite{schuricht,Dora2,Coll15}. Quantum quenches in LL have been studied in \cite{caz06,caz09,caz10,Dora,dbm} and a detailed comparison of our results with earlier works is presented in the Supplemental Material. 

Note that our method \cite{SotiriadisCalabrese,SM15} for the derivation of quench dynamics has certain advantages in comparison with others. By exploiting the connection between initial and evolved fields, we link the large time asymptotics of correlations directly to the long distance asymptotics of initial correlations, which are correctly described by Renormalization Group theory. Even though the applicability of this approach seems to be restricted to models with Gaussian evolution, in fact it is based on fundamental properties of quantum field theory that are also valid in the presence of interactions, more precisely the existence of a suitable set of local fields whose time evolution can be expressed as a convolution of initial local fields. 

In particular the conclusion that the GGE fails in the case of a genuinely interacting to free massless quench in $1d$ relies on the ballistic form of the evolution (\ref{dAlembert}) which is linked to the gaplessness of the energy spectrum: if the post-quench Hamiltonian was gapped instead, the Green's function in (\ref{eq:phi(t)}) would decay with the distance inside the light-cone and the quench dynamics would lead to the GGE \cite{SotiriadisCalabrese}. Therefore our result is expected to be robust under the insertion of irrelevant interactions to the post-quench Hamiltonian, as long as they do not affect the ballistic form of the evolution. Such perturbations of the Luttinger model preserve the gaplessness of the spectrum, even though they modify the dispersion relation from linear to nonlinear \cite{nonlinear}. The same is also expected to be true for quenches in $1d$ CFT in which there always exist local operators whose evolution is governed by the wave equation (\ref{wave}) (e.g. the energy-momentum tensor).

\emph{Experimental implementation} -- Ironically, our findings about the failure of the GGE can be experimentally checked in the same way that the GGE was actually observed in \cite{expo}, simply by starting from a non-Gaussian initial state, i.e. the ground state of a genuinely interacting Hamiltonian. We will show that this is feasible using present technology in the same experimental setup \cite{exp2}. 

One-dimensional bosonic models can be realized in cold atom systems \cite{exp1a} and quantum quenches can be implemented by tuning the system's parameters \cite{exp1c}. The most suitable experimental setup for this purpose is a system of two coupled $1d$ gases of interacting bosons prepared by condensate splitting \cite{GAPD1,GAPD2,GAPD3,GAPD4,PD}. The low-energy physics of this system can be derived through bosonization \cite{Giamarchi,GAPD1,GAPD2,GAPD3,GAPD4,gauss-split,gauss-split1,gauss-split2,fg15}. The density $n$ and phase field $\theta$ of the antisymmetric modes are described by the sine-Gordon model, with the interaction $a$ controlled by the tunneling coupling $J$ between the two Bose gases. For $J=0$ the interaction vanishes and the sine-Gordon model reduces to the Luttinger model i.e. the system corresponds to free gapless phonons. In the opposite limit of large $J$, solitonic excitations between neighboring vacua are suppressed and the system is described by essentially free gapped excitations above one of the degenerate vacua. 
The rapid tunability of the parameters of the experimental system allows the implementation and efficient study of quantum quenches. The subsequent evolution of correlation functions is analyzed through time-resolved measurements of matter-wave interference patterns after time-of-flight expansion of the gas, averaged over many repetitions \cite{exp1d}. Such measurements provide direct data for the multi-point correlations of the local phase difference $\theta(x)$ between the two condensates. 

These techniques made recently possible the celebrated first experimental observation of a GGE \cite{expo}. More precisely it was demonstrated that more than one `effective temperatures' are required for the description of the long time steady state of the system, which is a clear sign of absence of thermalization and equilibration to some type of GGE. In the regime of parameters used in the experiment, the system is well-described both before and after the quench by the Bogoliubov-de Gennes equations, which provide a mean-field (that is Gaussian) approximation \cite{exp1c,gauss-split,gauss-split2,fg15}. In particular the post-quench Hamiltonian corresponds to the gapless Luttinger model. According to our findings, equilibration in this case is correctly described by the standard GGE since the initial state is also Gaussian. 

Deviations from the latter would however be manifested if the initial state were non-Gaussian. In fact measurements of CCFs precise enough to reveal non-Gaussian features are also feasible in the same experimental setup and have been performed in the ground state of the system for a wide range of parameter values \cite{exp2}. This suggests that an experimental realization of a quantum quench of $J$ from an intermediate value to zero would make the observation of such deviations possible. Indeed, provided that the effective low-energy field theory gives a faithful description of the experimental system and quench protocol, in this case the initial state would be the ground state of the genuinely interacting sine-Gordon model and would exhibit non-Gaussian correlations. The parameters $\lambda_n$ of (\ref{lambda_nG}) can then be calculated using the methods of Integrable Field Theory \cite{Smirnov,mussardo-SFT}. Numerical checks can be performed by means of the so-called Truncated Conformal Space Approach \cite{TCSA,TCSA2} after a careful analysis of finite size effects. Details will be given in a forthcoming publication. Note however that eqs.(\ref{lambda_nG}), (\ref{v_nG}) and (\ref{vv_nG}) are more suitable for an experimental check: while it has been difficult to compare experimental measurements to theoretical predictions for the sine-Gordon model, it is quite easy to directly compare measurements on the initial ground state and at large-times after the quench.

\emph{Conclusions} -- 
We studied a quantum quench from an interacting to a free massless bosonic field theory in $1d$, deriving analytical results for the large time correlations and demonstrating that, in contrast to the massive case, the standard Gaussian GGE fails: the system retains memory of all initial correlations, which is accessible by local measurements. 

\

\acknowledgments
\emph{Acknowledgments} -- I am grateful to A. Bastianello for drawing my attention to the cumulant expansion, P. Calabrese, M. A. Rajabpour and M. Cramer for useful discussions. I acknowledge the ERC for financial support under Starting Grant 279391 EDEQS. The final part of this work has been carried out thanks to the support of the A*MIDEX project Hypathie (no. ANR-11-IDEX-0001-02) funded by the ``Investissements d'Avenir'' French Government program, managed by the French National Research Agency (ANR).

\section{SUPPLEMENTAL MATERIAL}

\subsection{Comparison with earlier work}

Quantum quenches in the context of Luttinger liquids have been studied in \cite{caz06,caz09,caz10,schuricht,Dora,Dora2,dbm,Coll15} and as a special case of CFT in \cite{cc06,cardy12}. Even though, as already mentioned, LL theory is only a low-energy description of $1d$ gapless systems, there is evidence that it captures correctly the quench dynamics in several cases \cite{schuricht,Dora2,Coll15,LLnew1,LLnew2} and certainly whenever high-energy excitations are suppressed, like when the quench is not performed instantaneously but within a short time interval. All results of \cite{cc06,caz06,caz09,caz10} are consistent with relaxation to the GGE, while the more general initial states considered in \cite{cardy12} lead to equilibration described by a GGE in which non-local charges are also expected to be included. The initial states considered in \cite{cc06,caz06,caz09,caz10}, when specialized to our settings, belong to the class of Gaussian initial states, therefore these results are in agreement with ours. Indeed the Dirichlet initial state used in \cite{cc06} exhibits vanishingly small correlation length $\xi_0\sim m_0^{-1}\to 0$ in which limit it is Gaussian. In this limit, from (\ref{lambda2}) we obtain $\lambda_2\sim m_0/2$ recovering the results of \cite{cc06}. 

On the other hand, all quantum quenches in the Luttinger model studied in \cite{caz06,caz09} correspond, after applying the bosonization mapping, to quenches within a Gaussian model. The same is true for the quench studied in \cite{caz10} since the Luther-Emery point of the sine-Gordon model is essentially free. The initial state considered in \cite{dbm} instead is a special type of non-Gaussian state and it would be interesting to check if relaxation retains memory of its non-Gaussianity, which is not however addressed in that work as the authors compare only with the thermal ensemble predictions. Lastly, the general quenches studied in \cite{cardy12} using CFT include also non-Gaussian perturbations in the initial state. The observation that memory of each of them survives at infinite times and must be included in the GGE in order for it to be correct, is consistent with our result for failure of the conventional GGE. 

In the literature an equivalent form of the Diagonal Ensemble (\ref{DE}) has appeared in a different context, the Lieb-Liniger model, and has been termed Generalized GGE \cite{ANDREIc}. However, while equilibration of a general initial state is always described by the Diagonal Ensemble \cite{gge2}, quench initial states, being ground states of local Hamiltonians, are special and so far there is no other known case of a quantum quench for which this ensemble does not reduce to some version of the much more economic GGE \cite{EPC-nonlocal}.
 
In lattice models, it has been proven in \cite{cdeo08,cra10} that a general initial state satisfying mild clustering requirements (as quench initial states do) and evolving under a gapless Gaussian Hamiltonian relaxes locally to a Gaussian reduced density matrix. However in that context local relaxation refers to single site observables \cite{cdeo08}, while for blocks of neighbouring sites \cite{cra10} the deviations from Gaussianity increase with the block size, meaning that in the scaling limit where the lattice spacing is taken to zero, correlations at finite distances are non-Gaussian. Consequently there is no inconsistency with our results. Extending those results to free fermion lattice systems, it was more recently shown \cite{eisert16} that, apart from clustering initial correlations, a second condition for Gaussian relaxation is that the evolution exhibits sufficiently delocalizing transport. This condition is shown to be typical in lattice systems, but is obviously not satisfied in the continuous bosonic system studied here, which exhibits instead ballistic transport.

An interesting question is whether the so-called Quench Action method \cite{ce-13,WDNBFRC,PMWKZT,bse-14,QA2} would give correct predictions in the cases considered here. This method approaches the problem of quench dynamics in a way different than the GGE and aims at describing the thermodynamic properties of the initial state of Bethe Ansatz solvable systems in an efficient and faithful way. For this reason and although there are no analytical calculations of steady state multi-point correlation functions so far, it is not expected that this approach would fail.

\end{document}